\documentclass[aps,physrev,reprint,groupedaddress]{revtex4-2}
\usepackage{graphicx}
\usepackage{hyperref}
\usepackage{color,soul}
\usepackage{amsmath}

\newcommand\lso{$\rm \alpha\text{-}Li_{2}SO_{4}$}
\newcommand{\Dtr}{D_{\mathrm{tr}}^{\mathrm{Li}}}
\newcommand{\Drot}{D_{\mathrm{rot}}^{\mathrm{SO}_{4}}}
\newcommand\SO{$\rm SO_{4}^{2-}$}
\newcommand\Li{$\rm Li^{+}$}

\begin{document}

\title{Coupling of Cation and Anion Dynamics in Solid Electrolytes Revealed by Virtual Isotopic Substitution Molecular Dynamics
}

\author{Xintian Wang}
\author{Andrew L. Goodwin}%
\author{Theodosios Famprikis}%
 \email{Contact author: theodosios.famprikis@chem.ox.ac.uk}
\affiliation{%
 Inorganic Chemistry Laboratory, University of Oxford, Oxford, United Kingdom
}%

\date{\today}

\begin{abstract}
We characterise the interplay between cation-translational and anion-rotational dynamics on the model system \lso\ through virtual isotopic substitution molecular dynamics (VISMD) simulations. We independently control in turn the lithium translational- and sufate rotational- diffusivity by varying the masses of lithium and oxygen, respectively, showing that both diffusivities exhibit power-law relationships with the masses of each individual component. Changing either diffusivity produces an effect on the other; \emph{i.e.} there exists a bidirectional coupling between anion and cation diffusive dynamics. From our variable-temperature/variable-mass dataset we demonstrate that the apparent activation energy for lithium diffusion depends strongly on the isotope mass, decreasing with accelerated sulfate dynamics; an observation which provides clear evidence of mode-coupling between cation and anion dynamics.
\end{abstract}

\maketitle

\section{Introduction}

Solid electrolytes are a class of materials in which a mobile ion can diffuse through a host framework, giving rise to strong translational diffusivity, $D_{\textrm{tr}}$, and macroscopic ionic conductivity. Although much about diffusion in the solid state can be inferred from the static, average host structure \cite{Catlow1983}, it has also long been recognised that the dynamics of the host framework can also play a key role through coupling of the vibrational or diffusive modes of the host to the diffusive modes of the mobile ion \cite{Bachman2016a,Muy2020, Wood2021}. In this context, a situation of particular interest involves cases where rotational tumbling of complex ions may influence the translational diffusion of separate atomic ions---often termed the \textit{``paddle-wheel''} mechanism~\cite{Zhang2022a}.

A case of extreme rotational dynamics is presented in so-called rotor-phase electrolytes, which exhibit at once plastically crystalline (\textit{i.e. rotor-phase}) and ion-conducting (\textit{i.e. electrolyte}) characteristics~\cite{Price_2003}. Plastic crystals are characterized by dynamic rotational disorder of a molecular or molecular-ionic component, which can be quantified by a rotational diffusivity, $D_{\textrm{rot}}$ \cite{Bee1988QENS}. The interplay between translational and rotational diffusivity in the same phase has been the subject of controversy in the literature~\cite{Lunden1988, Jansen1991a,Andersen1992, Jun2024, Smith2025}; here we occupy ourselves with describing the coupling between $D_{\textrm{rot}}$ and $D_{\textrm{tr}}$ in rotor-phase electrolytes, as a model case for the wider question of coupling between host and mobile ion dynamics in solid electrolytes.

The high-temperature phase of lithium sulfate, \lso, is an archetypical rotor-phase electrolyte, combining rotational diffusivity of the sulfate molecular anion and translational diffusivity of the lithium cation. The phase transition from the low-temperature $\beta$-phase at 850\,K brings about the onset of \SO-rotational diffusivity and a stark increase in \Li-translational diffusivity in \lso~\cite{Kvist1973, Nilsson_80, Aronsson1983}. Such apparent coupling of the two diffusive modes through the phase transition to the plastically crystalline phase is also observed for a host of related molecular-ionic inorganic compounds, such as $\rm\alpha\text{-}Na_{3}PO_{4}$~\cite{Witschas2000b}, $\rm\gamma\text{-}Na_{3}PS_{4}$~\cite{Famprikis2019,Sau2020}, $\rm\gamma\text{-}Na_{4}P_{2}S_{6}$~\cite{Scholz2022, Hogrefe2025},
$\rm Li_2OHCl$~\cite{Dawson2018,Wang2020},
a variety of lithium and sodium borohydrides \cite{DeJongh2016,Cerny2022,Liu2023} \emph{etc.}---as well as related organic-ionic plastic crystal electrolytes \cite{Pringle2010,Thomas2023}.

Historically, discussions of diffusive-translation--rotation coupling in such rotor-phase ion conductors have been conducted through the lens of the so-called \textit{``paddle-wheel''} mechanism~\cite{Kvist1973}. This mechanism has always been appealing from an intuitive perspective, but its scientific veracity continues to spark controversy~\cite{Lunden1988, Jansen1991a,Andersen1992, Jun2024, Smith2025}. One difficulty is that various overlapping interpretations of the metaphor exist as already highlighted early on in Ref.~\citenum{Jansen1991a}: \emph{i.e.} active transport of the cations by cooperative rotational movement of the anions, and/or independent anion rotations enhancing the cation mobility through transient bottleneck opening, and/or anions being free to rotate out of the way in response to cation hopping attempts and/or the anion rotational diffusion favoring cation diffusion through periodically modulating the cation energy landscape. This lack of clarity risks confusing the debate and hindering our understanding of the dynamical interplay in rotor-phase electrolytes.

Furthermore, much of the \textit{paddle-wheel} debate has focused on how anion rotations might facilitate cation translation~\cite{Zhang2022a}; in other words, it is implied that ion diffusivity follows causally from anion rotation (\textit{linear causality}). This bias is likely influenced by the technological relevance of ion conduction as a materials property, and the fact that rotor-phase electrolytes have mostly been studied in the context of their potential for electrochemical applications \cite{Zhang2022a}. We will come to show that there exists no causal directionality in the coupling: cation and anion dynamics mutually come about as a result of one another; in other words, this is a \textit{chicken--egg} situation (\textit{circular causality}).

Our approach is to use molecular dynamics simulations of the model system \lso, where we independently vary the rotational and translational dynamics to uncover the relationship between rotational and translational diffusivities. We achieve this using a virtual isotopic substitution molecular dynamics (VISMD) approach to vary diffusion rates without affecting the physical or chemical driving forces at play. In doing so, we uncover a reciprocal correlation between rotational and translational diffusivity; that is the two quantities scale with one another regardless of direction. This implies a \textit{reciprocal causation}, where accelerated lithium-ion diffusion causes accelerated sulfate-ion reorientation and \emph{vice versa}. This key result points to a synergistic, intrinsically coupled view of the two types of dynamics that refines our understanding of \lso\ and likely applies to the range of known rotor-phase electrolytes. We expect that this insight will critically inform further discussions of the anion-cation dynamic coupling in solid electrolytes and decisively accelerate their theoretical and applied development.

\section{Results and Discussion}
\subsection{Molecular Dynamics of \lso}
\begin{figure}[tb] 
\includegraphics[width=0.9\columnwidth]{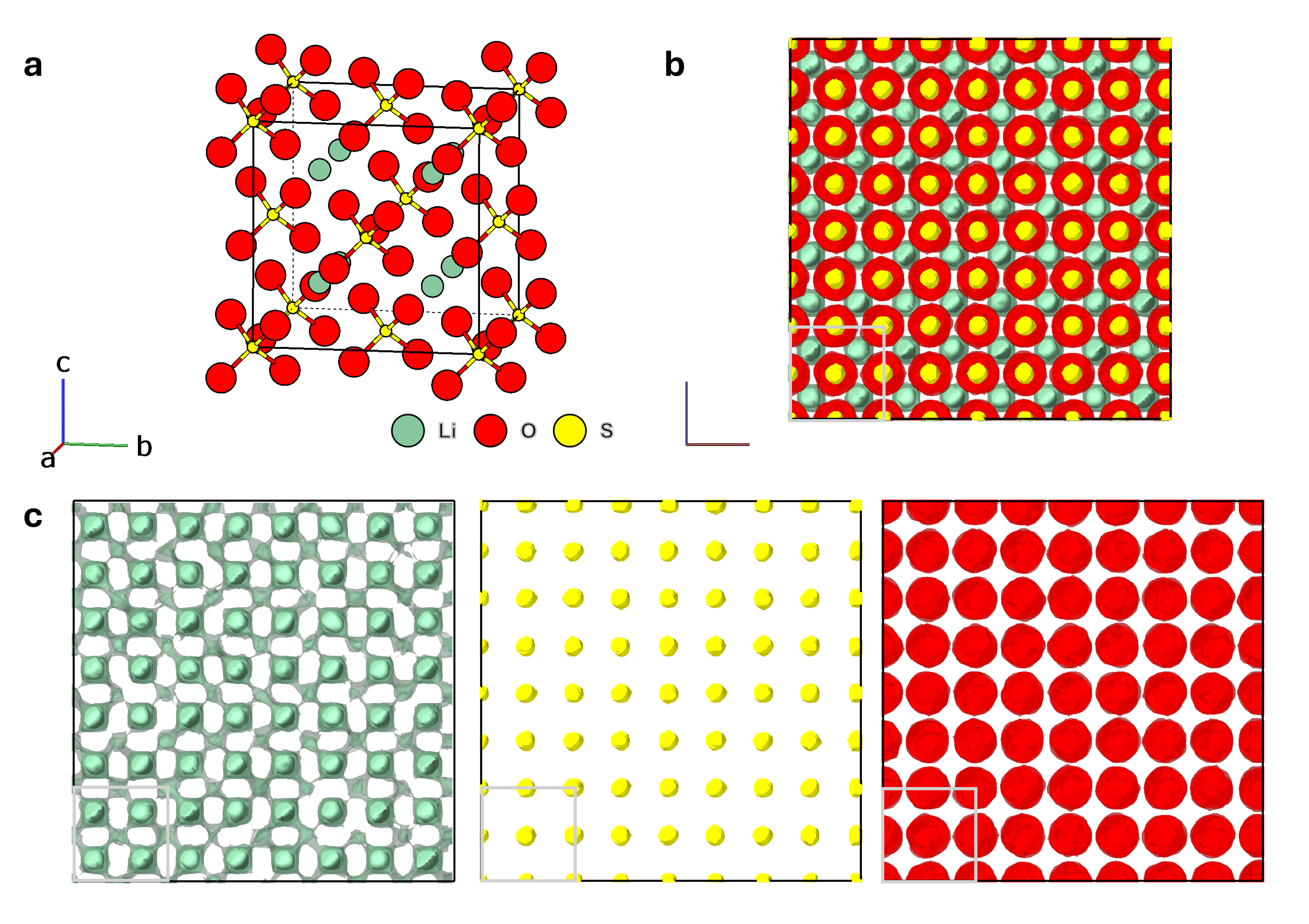}
\caption{Average structure of \lso. (a) Approximate crystallographic unit cell based on Ref.~\citenum{Forland1957}. (b,c) Probability density coloured per atom from a molecular dynamics simulation of a $4 \times 4 \times 4$ supercell of \lso\ at 1000\,K}
\label{MD_densities}
\end{figure}

Figure~\ref{MD_densities} exemplifies the structure and dynamics of \lso\ as probed using molecular dynamics simulations at 1000\,K. \lso\ adopts an antifluorite-like crystal structure with sulfate anions arranged in a face-centered-cubic arrangement and lithium cations occupying preferentially the corresponding tetrahedral interstitials, consistent with previous experimental and computational results \cite{Forland1957,Nilsson_80,Impey1985, Kaber1992, Karlsson1995}. Lithium ion translational diffusion proceeds through hopping between tetrahedral and octahedral sites; likewise, sulfate ion rotational diffusivity is evident from the smearing of the time-averaged oxygen-atom probability density in shells around the sulfur atoms.

We quantify the lithium-ion translational diffusivity from the linear evolution of the lithium mean-squared displacement \cite{Mehrer2007}
\begin{equation}
    \Dtr =
    \frac{1}{6}
    \frac{\mathrm{d}}{\mathrm{d}\Delta t}
    \left\langle
        \left|
            \mathbf{r}_{\mathrm{Li}}(t + \Delta t)
            - \mathbf{r}_{\mathrm{Li}}(t)
        \right|^2
    \right\rangle,\label{MSD}
\end{equation}
and the sulfate-ion rotational diffusivity from the characteristic time-decay of the S--O bond orientational correlation function \cite{Burshtein1994}
\begin{equation}
    \Drot=-\frac{1}{2}\frac{\rm d}{{\rm d}\Delta t}\ln\left\langle
\mathbf{n}_{\mathrm{S}\text{-}\mathrm{O}}(t+\Delta t)
\cdot
\mathbf{n}_{\mathrm{S}\text{-}\mathrm{O}}(t)
\right\rangle.\label{acf}
\end{equation}
Here $t$ denotes simulation time, $\mathbf{r}_{\mathrm{Li}}$ are lithium position vectors, $\mathbf{n}_{\mathrm{S}\text{-}\mathrm{O}}$ are S-O bond vectors and brackets denote averages over all equivalent time segments $\Delta t$ and all lithium atoms or sulfur--oxygen bonds, respectively.

Our VISMD method for investigating the correlation between lithium-translational and sulfate-rotational diffusivity relies on carrying out molecular dynamics simulations with varying masses of lithium or oxygen. At a given temperature (\textit{i.e.} average kinetic energy), scaling a component's mass will scale its velocity. Thus, we expect that changing the mass of lithium scales $\Dtr$, while altering the mass of oxygen changes the moment of inertia of \SO and scales \(\Drot\). Crucially, holding the potential parameters constant ensures all static properties of the system remain unchanged, allowing for meaningful investigation of the interplay between the diffusive modes. 

Analogous approaches have been used before to study the effect of particle mass on the dynamic properties of liquids \cite{Bearman1981,Nuevo1995, Kiriushcheva2005} and ionic melts \cite{Liu2018}. In the solid-electrolyte literature, simulations are sometimes performed with the framework atoms fixed (\textit{i.e.} equivalent to infinite isotopic mass) to infer dynamic correlation effects (\textit{e.g.} refs. \cite{Kahle2018,Singh2024, Wang2020}). A few molecular dynamics studies on rotor-phase electrolytes have included virtual isotopic substitution \cite{Ferrario1995,Yin2004,Oh2025}, but systematic variable-mass molecular dynamics simulations are not commonplace in the wider solid-electrolyte literature. Still, the effect of atomic mass on diffusion in solids has been treated both theoretically and experimentally (incl. studies on $\rm^{6/7}Li_2SO_4$ \cite{Kvist1966}), in which context it is termed the \textit{``isotope effect''} or \textit{``mass effect''} \cite{Mehrer2007,Jing2024}.

\begin{figure}[tb] 
\includegraphics[width=0.9\columnwidth]{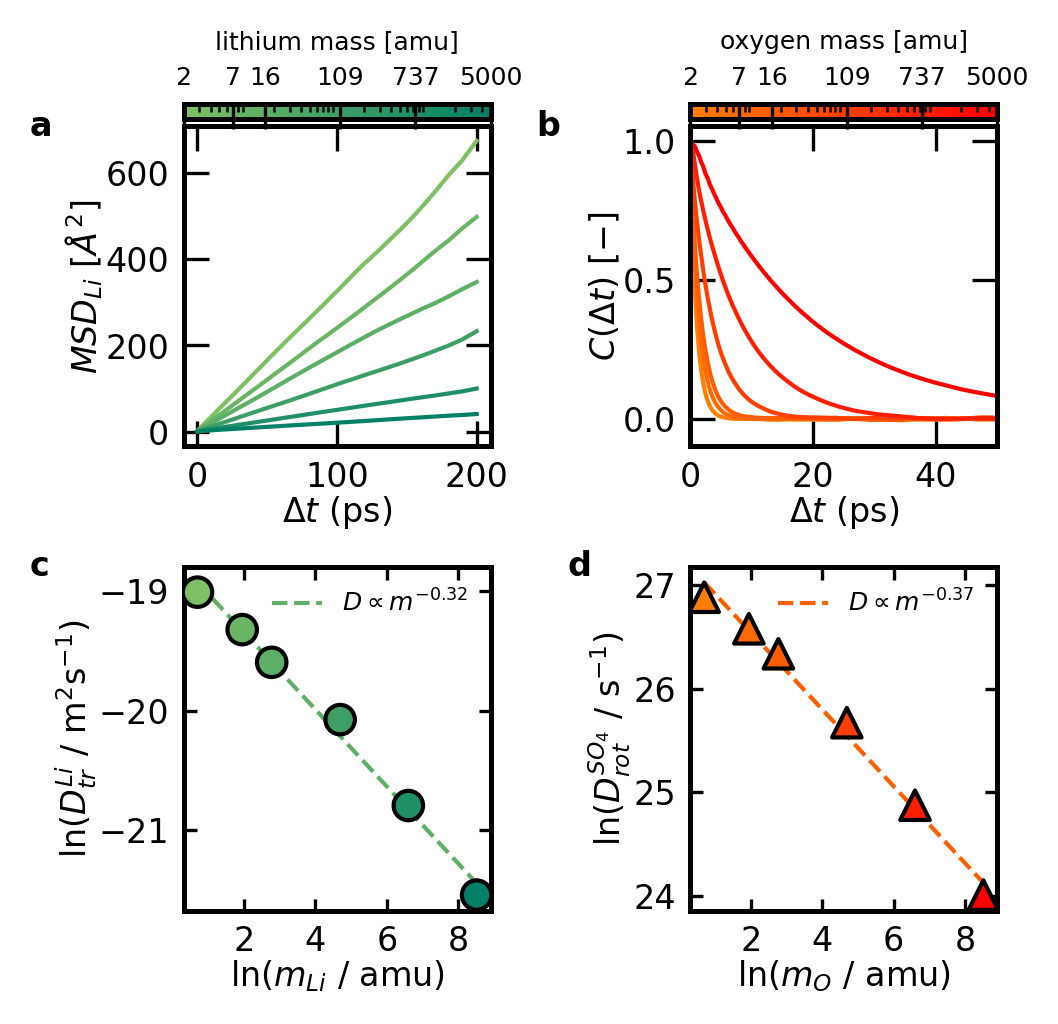}
\caption{Independent control of rotational and translational dynamics in \lso\ at 1000\,K. (a) Mean-squared-displacement of lithium as a function of its mass. b) S--O bond self-correlation as a function of oxygen mass c) Lithium-ion translational diffusivity as a function of lithium mass and d) sulfate-ion rotational diffusivity as a function of oxygen mass.}
\label{D_m_homo}
\end{figure}

Figure \ref{D_m_homo} shows that the lithium-translational and sulfate-rotational diffusivities calculated from our simulations vary as expected (\textit{i.e.} inversely) with the lithium and oxygen mass, respectively. To a good first approximation this scaling obeys simple power-law relationships of the form $D=am^{b}$. We extract the exponents $b$ through linear fits of the log--log plots of diffusivity versus mass, determining values of $-0.32$ and $-0.37$ for the lithium and oxygen cases, respectively, at 1000\,K. We note that in the classical limit for a single ion diffusing in a solid  one would expect a value of $-0.5$ for the exponent; \textit{i.e.} $D\propto \nu_{0} \propto 1/\sqrt{m}$, for a characteristic attempt frequency $\nu_0$ \cite{Jing2024}.

\begin{figure}[tb] 
\includegraphics[width=0.9\columnwidth]{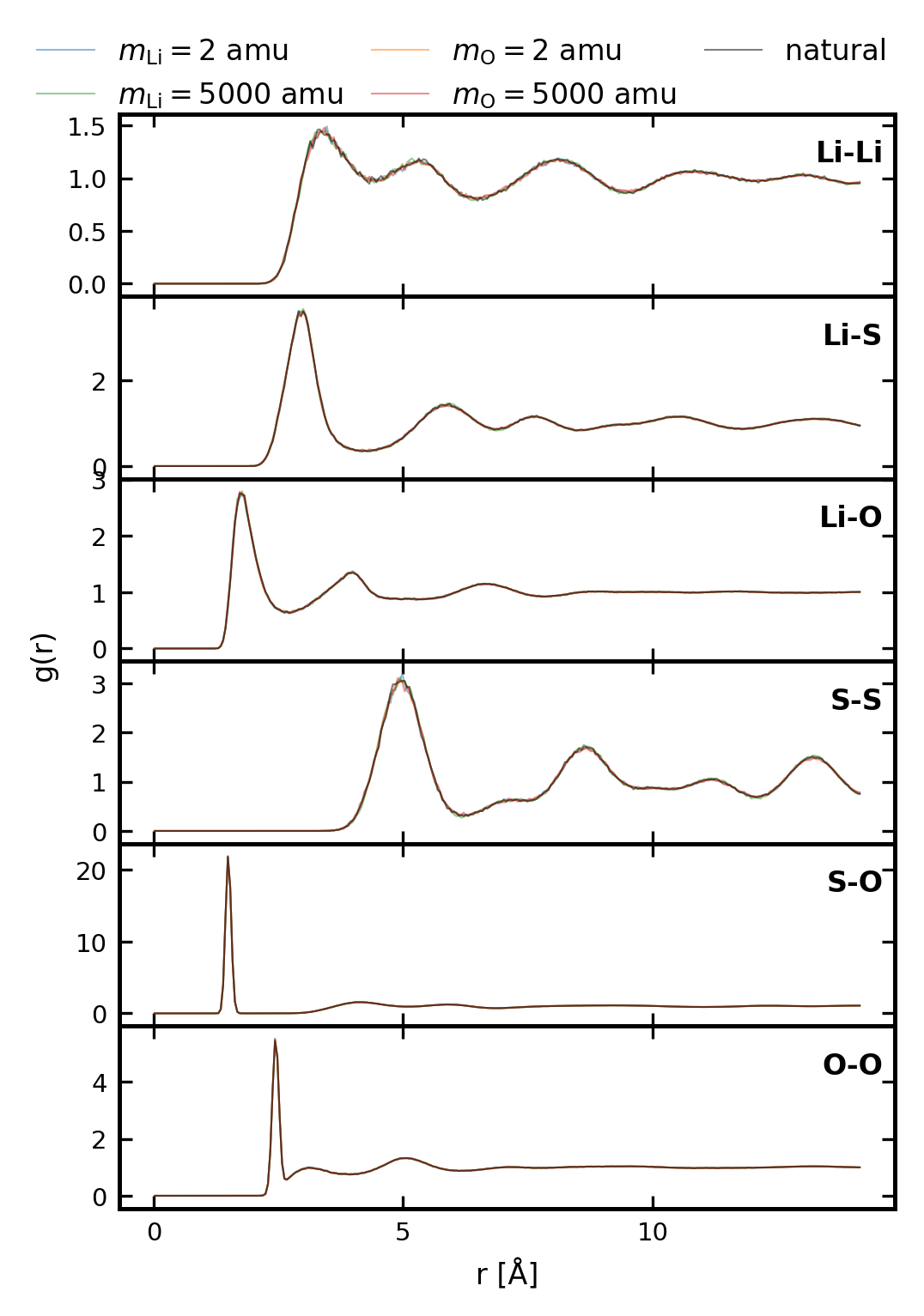}
\caption{Partial pair distribution functions at 1000\,K with natural, reduced and increased lithium and oxygen masses.}
\label{RDF_vs_mass}
\end{figure}

Figure~\ref{RDF_vs_mass} shows the radial distribution functions calculated from our VISMD simulations with the extreme (highest and lowest) masses of lithium and oxygen; it is evident that the isotopic substitution has no effect on the average structure of the system as the corresponding distribution functions are indistinguishable. Hence, this approach allows us to independently vary the magnitude of the translational or rotational diffusion coefficients, while controlling for the chemical interactions and average structure of \lso. 

\subsection{Translation--Rotation Reciprocal Coupling}

\begin{figure}[tb] 
\includegraphics[width=0.9\columnwidth]{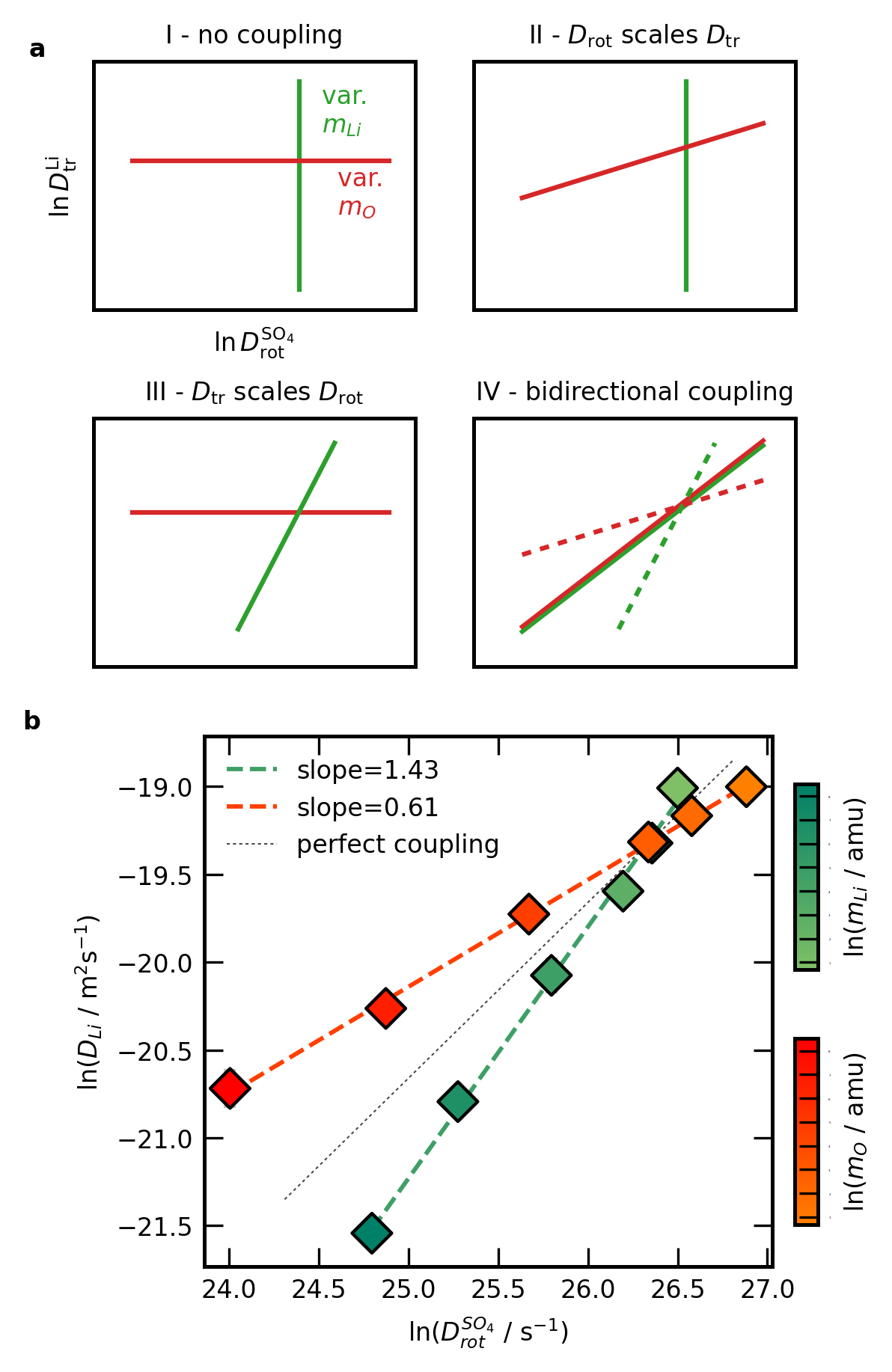}
\caption{(a) Possible scenarios of diffusive-translation--rotation coupling in \lso. (b) Correlation between lithium-translational and sulfate-rotational diffusivities as they vary through lithium mass and oxygen mass at 1000 K.}
\label{Drot_Dtr}
\end{figure}

Our next step was to explore the degree of coupling between Li$^{+}$-ion translations and SO$_4^{2-}$-ion rotations. Anticipating the various possible scenarios, we consider qualitatively the possible limiting cases of the relationship between $\Dtr$ and $\Drot$ schematically shown in Figure ~\ref{Drot_Dtr}(a). Each panel in that Figure shows two lines corresponding to varying the masses of one of the two components (Li or O), and the two lines always intersect at the points corresponding to natural isotopic masses. Scenario I presents the expected results in the case of no coupling: changing the mass of Li only scales $\Dtr$ and leaves $\Drot$ unaffected; likewise changing the mass of oxygen only scales $\Drot$ and not $\Dtr$. Scenarios II and III involve unidirectional correlations. Specifically, scenario II represents one possible interpretation of the \textit{paddle-wheel} hypothesis, whereby changing the mass of lithium only affects $\Dtr$ but changing the mass of oxygen controls not only $\Drot$ but also $\Dtr$ through coupling; \textit{i.e.} the cation diffusivity is facilitated by anion rotation. Scenario III is the inverse case of II, where $\Drot$ is coupled to $\Dtr$ but not \emph{vice versa}. Scenario IV presents the case of bidirectional coupling: varying either $\Dtr$ or $\Drot$ produces changes in the other. In the limit of perfect coupling, the lines for varying lithium- and oxygen mass would converge, producing a plot with unity slope.

Figure ~\ref{Drot_Dtr}(b) shows the various translational and rotational diffusion coefficients extracted from our VISMD simulations at 1000\,K for a range of lithium and oxygen masses. Our data clearly show that changing either the lithium or oxygen mass produces linear changes in both diffusion coefficients. This is clear evidence of bidirectional coupling of the two phenomena, corresponding to scenario IV. Indeed, we find that $\Dtr$ scales with the oxygen mass, and $\Drot$ scales with the lithium mass according power laws similar to those shown in Fig. \ref{D_m_homo}. This observation implies that the ionic interactions between $\rm Li^{+}$ and $\rm SO_{4}^{2-}$ impose a coupling of their respective dynamics; that is rotational motion of sulfate units is no more responsible for lithium-ion transport than lithium-ion displacements are for rotational diffusion of the sulfate.

Mode-coupling theory provides the appropriate formalism within which to describe the observed interplay between translational and rotational diffusion in rotor-phase electrolytes \cite{Gotze2008,Das2004,Janssen2018}. Coupling between rotational and translational diffusion has already been discussed in the context of viscous liquids~\cite{Bagchi2001}, where, in the perfect coupling limit, the rotational diffusivity of a solute has been shown to be proportional to the translational diffusivity of the surrounding solvent molecules, with
\begin{equation}
    \frac{D_{\mathrm{tr}}}{D_{\mathrm{rot}}} 
    =
    \mathrm{constant}.
\end{equation}
It follows that 
\begin{equation}
\frac{\mathrm{d}\ln\left(D_{\mathrm{tr}}\right)}
         {\mathrm{d}\ln\left(D_{\mathrm{rot}}\right)}
    = 1
\label{pc}
\end{equation}
in this regime of perfect coupling (scenario IV in Figure \ref{Drot_Dtr}(a).

In the case of \lso, \SO\ acts as the reorienting solute molecule and \Li\ the translationally diffusing solvent. The higher the lithium diffusivity, the lower the effective ``solvent viscosity'' and thus the higher the \SO\ rotational diffusivity. The argument can be reversed by considering \Li\ as the solute in a medium whose viscosity is inversely proportional to the \SO\ rotational diffusivity, constituting the bidirectionality of the correlation observed.

Our data for \lso\ give
\begin{equation}
\frac{\mathrm{d}\ln\left(D_{\mathrm{tr}}\right)}
         {\mathrm{d}\ln\left(D_{\mathrm{rot}}\right)}
    =0.61
\end{equation} 
for the variable-oxygen-mass series and 
\begin{equation}
\frac{\mathrm{d}\ln\left(D_{\mathrm{tr}}\right)}
         {\mathrm{d}\ln\left(D_{\mathrm{rot}}\right)}
    = 1.43
\end{equation}
for the variable-lithium-mass series, each at 1000\,K. The deviation of these slopes from unity is a direct measure of the real variation away from ideal coupling. So we find that \lso\ is actually intermediate to scenario I (no coupling) and the ideal limiting case of scenario IV in Fig.~\ref{Drot_Dtr}(a) (eq.~\eqref{pc}). The degree of coupling at 1000\,K is about 60\%. We note that the slopes for the two datasets are approximately reciprocal, which highlights the strong bidirectionality of the correlation. 

\subsection{Coupling Evolution with Temperature}

\begin{figure}
\includegraphics[width=0.9\columnwidth]{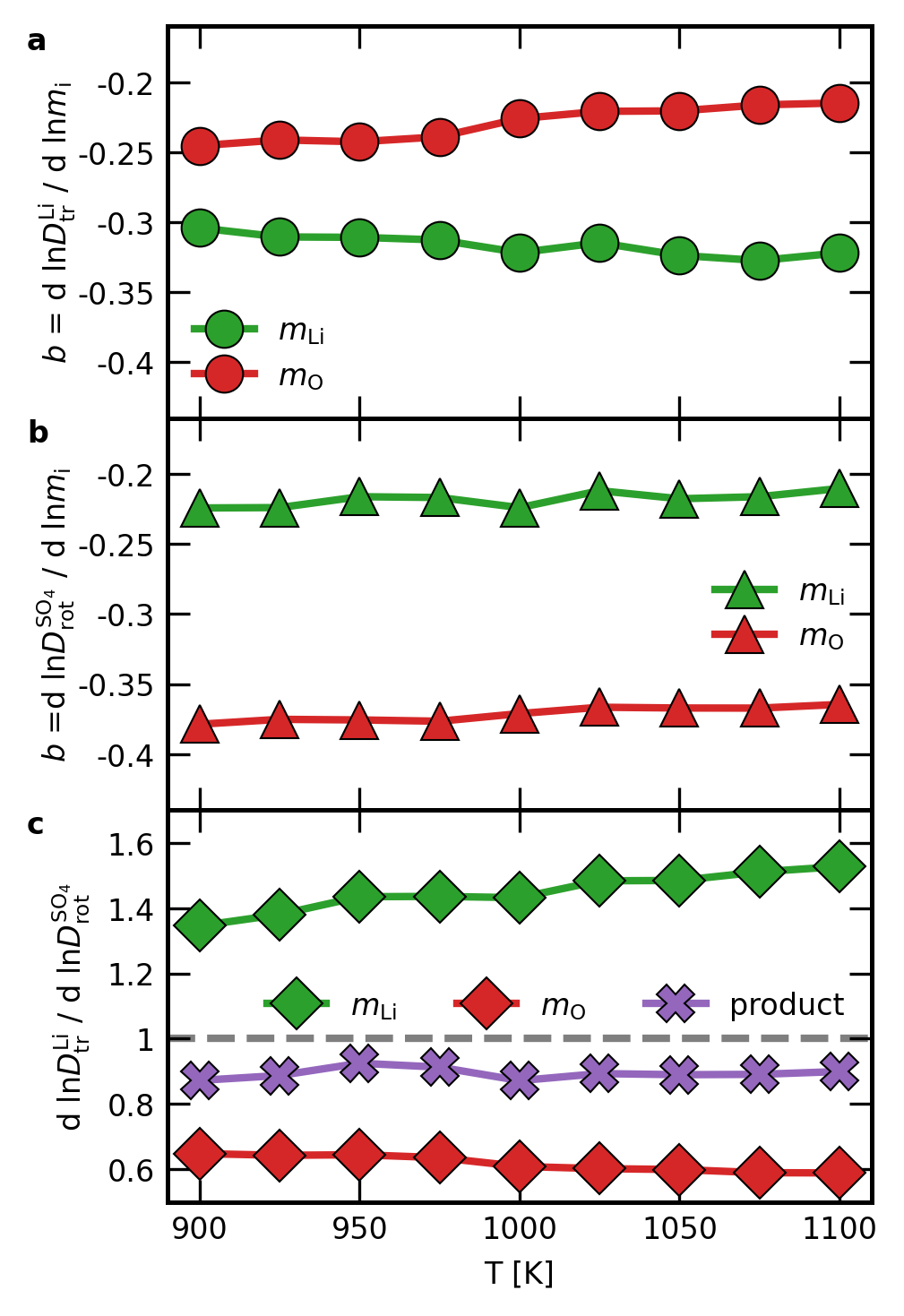}
\caption{Evolution of the rotation-translation coupling with temperature.  Power-law exponents of (a) lithium translational diffusivity and (b) sulfate rotational diffusivity; as a function of the mass lithium (green) and oxygen (red). (c) Corresponding translation-rotation coupling strengths}
\label{slopes_T}
\end{figure}

The diffusivity-mass power-law exponents
\begin{equation*}
    b=\frac{{\rm d}\ln(D_{i})}{{\rm d}\ln(m_{i})}
\end{equation*}
and coupling strengths
\begin{equation*}
\frac{{\rm d}\ln(D_{\textrm{tr}})}{{\rm d}\ln(D_{\textrm{rot}})}
\end{equation*}
for the Li- and O- isotopic simulation series evolve with temperature (Fig.~\ref{slopes_T}). We find that the coupling strength decreases on heating, which is likely a simple effect of increased anharmonicity. Nevertheless, the product of the slopes [Fig.~\ref{slopes_T}(c)] serves as a descriptor of the bidirectionality of the coupling, and always remains constant $\sim1$ throughout the temperature range simulated.

\begin{figure}[tb]
\includegraphics[width=0.9\columnwidth]{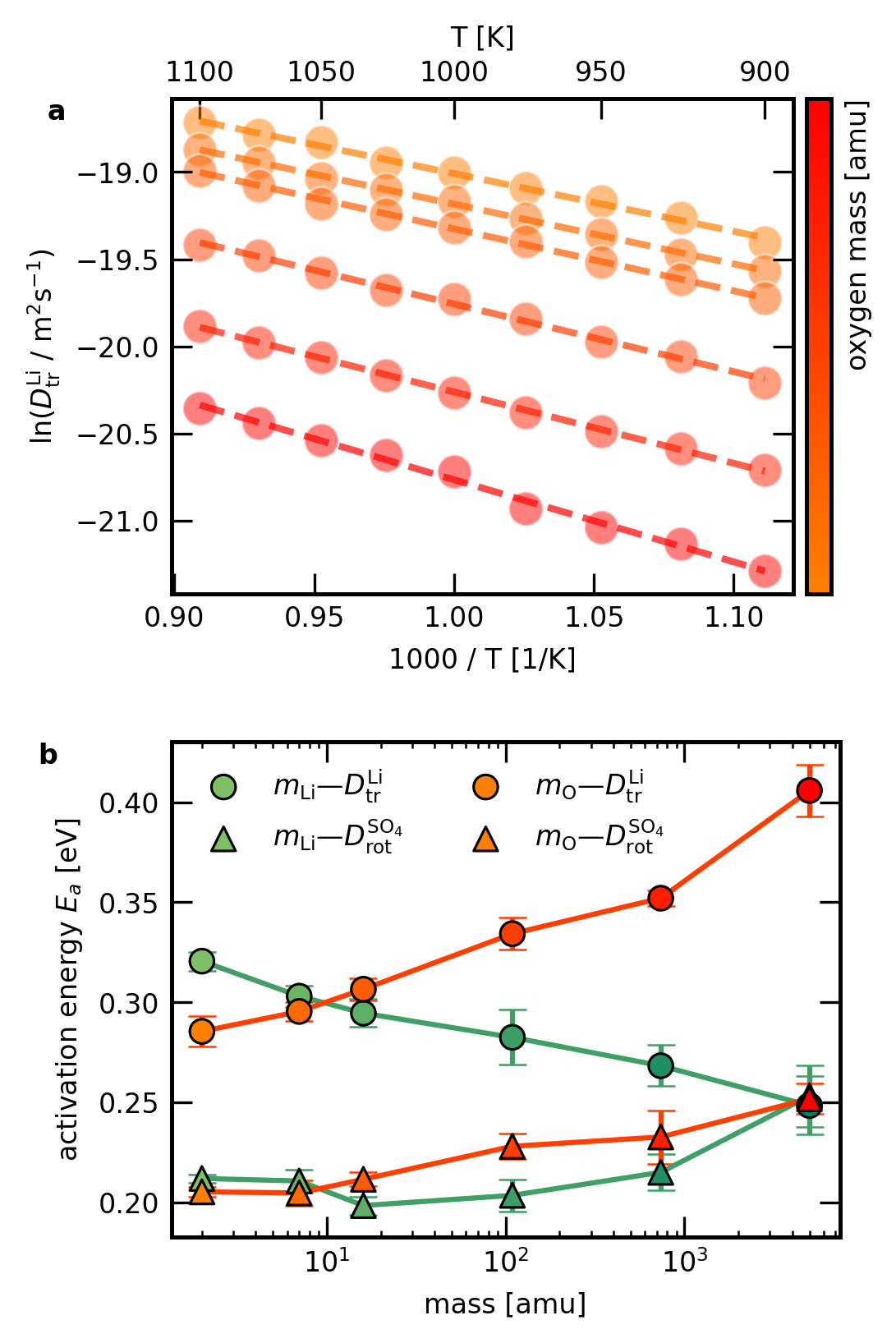}
\caption{(a) Lithium diffusivity as a function of oxygen mass and temperature (b)Evolution of activation energy for $\Dtr$ (circles) or $\Drot$(triangles) as a function of the mass of lithium (green) and oxygen (red)}.
\label{Ea_vs_mass}
\end{figure}

Another instructive way to analyse the temperature evolution of the coupling is to extract the temperature dependence of $\Dtr$ and $\Drot$ (expressed as activation energies, $E_{a}$) as a function of mass; we do precisely this in Fig.~\ref{Ea_vs_mass}. Panel (a) shows the example of the Arrhenius temperature evolution of $\Dtr$ for the variable-oxygen mass series, and panel (b) collates the corresponding results for both diffusivities and both variable-mass simulation series.

Under the assumption that the activation energy reflects the free-energy barrier for an elementary diffusion event (i.e. \Li\ hop or \SO\ reorientation), diffusivity scales with temperature as
\begin{equation}
D(m,T) = D_0(m) \exp\left(-\frac{E_{\rm a}}{k_B T}\right).
\end{equation}
Changing the mass of components ought only to scale the (temperature-independent) prefactor $D_0(m)$ while leaving the energy landscape (and thus the activation energy) unaffected. This is indeed what occurs for $\Drot$, whose activation energy varies only very weakly in our simulations with varying mass of lithium or oxygen [Fig.~\ref{Ea_vs_mass}(b)]. In other words, the sulfate reorientation activation energy is---to a first approximation---a static configurational barrier.

In contrast, Figure~\ref{Ea_vs_mass} reveals a strong dependence of the activation energy for $\Dtr$ as a function of component mass; namely the activation energy decreases with increasing Li mass, and increases with increasing O mass. This result is a clear indication that the apparent activation energy for lithium diffusion is not solely dependent on the static/average structure (which does not vary with mass, \emph{c.f.} Fig.~\ref{RDF_vs_mass}), but is strongly coupled to the dynamics. Changing the temperature evidently changes the degree of dynamical coupling (Fig.~\ref{slopes_T}) and this is reflected in a change in the apparent activation energy. In other words, the apparent activation energy can be thought of as the combination of the (mass independent) static free-energy barrier and a (mass dependent) dynamical-coupling component which can be positive or negative:
\begin{equation}
E_{\mathrm{a}}^{\mathrm{app}} \approx E_{\mathrm{static}}^{\mathrm{Li~hop}} + E_{\mathrm{coupling}}(m_{\mathrm{Li}},m_{\mathrm{O}}).
\end{equation}
We note that the trends shown in Figs.~\ref{slopes_T}(a,b) and \ref{Ea_vs_mass}(b) are phenomenologically related, such that for a given diffusivity and a given component mass the evolution of the activation energy with mass is proportional to the evolution of the power-law exponent $b$ with temperature (see Appendix). In particular,
\begin{equation}
\frac{1}{k_{\mathrm{B}}T^2}
\frac{\partial E_{\textrm a}}{\partial \ln(m)}
=
\frac{\partial b}{\partial T}.
\label{eq:b_Ea}
\end{equation}

The trend of increasing $E_{\textrm a}^{\textrm{Li}}$ with increasing oxygen mass (Fig.~\ref{Ea_vs_mass}) is of particular interest as it underlines the strong positive effect of accelerating sulfate dynamics on lithium-ion diffusion. This is consistent with the inference that anion rotational diffusivity decreases the activation energy for cation diffusivity, which had been proposed in early studies of rotor-phase electrolytes \cite{Jansen1991a}. Further nuanced simulations controlling for \SO\ displacement (through the mass of sulfur) and diffusion-event-based analysis would shed further light on the nature of the coupling mechanism.

A recent study on the rotor-phase electrolyte $\rm Na_{3}(B_{12}H_{12})(BH_{4})$ also probed the effect of anion dynamics on cation diffusion by isotopic molecular dynamics \cite{Oh2025}. Similar to our results, an inverse relationship of the sodium-ion conductivity and anion rotational speed with anion mass was observed, but with only a modest effect on the activation energy for ion conduction. This result indicates that even among rotor-phase electrolytes, the nature and magnitude of diffusive-rotational--translational coupling varies, albeit for reasons that remain to be determined.

In closing, we note that the interplay between cation and anion dynamics is not only of interest for the case of rotor-phase electrolytes, but in fact all solid electrolytes and especially those composed of molecular anions with additional rotational degrees of freedom. Coupling between anion dynamics and cation diffusion have been widely suggested in the literature on both experimental and computational grounds, \textit{e.g.} for  $\rm Li_3PS_4$ \cite{Zhang2020,Forrester2022}, $\rm Li_{10}GeP_{2}S_{12}$ \cite{Adams2012, Kahle2018}, $\rm NaMCl_6 (M=Nb,Ta)$ \cite{Li2025}, $\rm Li_{3}N$ \cite{Kahle2018}, AgI \cite{Brenner2020} etc. In light of the bidirectional correlation between cation- and anion-dynamics we put forth here, the observation of higher vibrational amplitudes in higher ionically conducting phases could just as well be the result (rather than the cause) of the increased cation dynamics; and we content that isotopic molecular dynamics could provide a convenient widely applicable probe to elucidate the underlying coupling mechanisms.

\section{Conclusion}
We believe the results presented above provide a conceptually important classification of the concept of translation--rotation coupling in molecular-ionic ion conductors. The pervasive ``paddle-wheel'' hypothesis has as its central inference that anion-rotational dynamics must affect cation-translational diffusion in solids. Instead we show here that for the model system \lso\ this is in fact a ``chicken-egg'' question, in that both diffusive modes come about concurrently upon the $\beta$-to-$\alpha$ phase transition and are intrinsically coupled to one another. Methodologically, we argue that performing molecular dynamics simulations with variable-masses is a simple, perhaps underused approach to derive profound insight into the nature of dynamically disordered solids.

\section{Methods}

Molecular dynamics were performed with the LAMMPS software package \cite{LAMMPS}. The empirical potentials used for $\rm Li_{2}SO_{4}$ where those of Parfitt \textit{et al.} \cite{Parfitt2005} building on earlier work by Klein \textit{et al.} \cite{Klein1983,Ferrario1995}.  Ionic interactions were described by a Coulomb(–Buckingham) potential
\begin{equation}
    v_{\alpha,\beta}(r)
= A_{\alpha,\beta}\exp\!\left(-a_{\alpha,\beta}r\right)
+ \frac{q_\alpha q_\beta e^2}{4\pi \epsilon_0 r},
\label{Buck}
\end{equation}
while intramolecular interactions in the sulfate anions were modeled by S--O and O--O harmonic terms\begin{equation}
v_{\alpha,\beta}(r)
= \frac{1}{2} k_{\alpha,\beta}
\left(r - R_{\alpha,\beta}\right)^2.
\label{harmo}
\end{equation}
Here $q$ denotes the atomic charge 
($q_{\mathrm{Li}}=+1.0$, $q_{\mathrm{S}}=+1.2$, $q_{\mathrm{O}}=-0.8$), $A$ and $a$ are the pre-exponential factor and repulsion parameter, $\epsilon_0$ is the  permittivity of free space, and $k$ and $R$ represent the force constant and equilibrium bond length, respectively. The full set of potential parameters taken from  Parfitt \textit{et al.} \cite{Parfitt2005} is listed in Table \ref{potential}.

\begin{table}
\centering
\caption{Potential parameters for \lso\ used in the molecular dynamics simulations.}
\begin{tabular}{ccc}
\hline\hline
\multicolumn{3}{l}{Exponential terms (eq.~\eqref{Buck})} \\
$\alpha$--$\beta$ & $A_{\alpha\beta}$ (MJ\,mol$^{-1}$) & $a_{\alpha\beta}$\,(\AA$^{-1}$) \\
\hline
O--O  & 234.0 & 4.180 \\
O--Li & 16.5  & 3.255 \\
\multicolumn{3}{l}{Harmonic terms (eq.~\eqref{harmo})} \\
$\alpha$--$\beta$ & $k_{\alpha\beta}$\,(MJ\,mol$^{-1}$\,AA$^{-2}$) & $R_{\alpha\beta}$\,(\AA) \\
\hline
S--O & 1.278 & 1.51 \\
O--O & 1.062 & 2.46 \\
\hline\hline
\end{tabular}
\label{potential}
\end{table}

Simulations were conducted
using a $4 \times 4 \times 4$ supercell of the high-temperature \lso\ cubic structure. The canonical ensemble ($NVT$) was simulated using a Nosé--Hoover thermostat. A cutoff radius of $9.7$\,\AA was applied for the pairwise interactions, and the Ewald summation method with a precision of $1 \times 10^{-4}$ was used for the long-range electrostatics. A timestep of $0.5$\,fs was employed, and the system was simulated for a total of $200$\,ps.

Simulations were performed at multiple temperatures in the range of 800 to 1400\,K and various lithium and oxygen masses in the range of 2 to 5000\,amu. At each temperature, the equilibrium volume was determined by simulations in the $NPT$ ensemble, and subsequently fixed for the $NVT$ production runs. Analysis of translational and rotational diffusion was performed using the GEMDAT python package \cite{Lavrinenko2026}.

\begin{acknowledgments}
T.F and A.L.G were generously supported by the Royal Society (Newton International Fellowship  \verb|NIF\R1\231784|, and Faraday Discovery Fellowship, respectively). A.L.G. gratefully acknowledges M.\,T.\,Dove (QMUL) for useful discussions. T.F gratefully acknowledges N.\,Benshalom (Oxford) for useful discussions.
\end{acknowledgments}

\bibliography{li2so4_alg}

\newpage

\appendix

\section{}

We consider two equivalent parameterizations of the diffusion coefficient. First, the Arrhenius form
\begin{equation}
D(T,m)
=
D_0(m)
\exp\left[(
-\frac{E_{\rm a}(m)}{k_{\mathrm{B}}T}
\right],
\label{eq:arrhenius}
\end{equation}
where both the prefactor $D_0$ and activation energy $E_{\rm a}$ may depend on the particle mass $m$. Empirically, the mass dependence at fixed temperature is well described by a power law:
\begin{equation}
D(T,m)
=
a(T)\times m^{b(T)}.
\label{eq:powerlaw}
\end{equation}
Equating Eqs.~\eqref{eq:arrhenius} and \eqref{eq:powerlaw} and taking the natural logarithm yields
\begin{equation}
\ln D_0(m)
-
\frac{E_{\rm a}(m)}{k_{\mathrm{B}}T}
=
\ln a(T)
+
b(T)
\ln m.
\label{eq:logrelation}
\end{equation}
Differentiating Eq.~\eqref{eq:logrelation} with respect to temperature at fixed mass then gives
\begin{equation}
\frac{E_{\rm a}(m)}{k_{\mathrm{B}}T^2}
=
\frac{\partial \ln a}{\partial T}
+
\ln m
\frac{\partial b}{\partial T}.
\label{eq:dT}
\end{equation}
Finally, differentiating Eq.~\eqref{eq:dT} with respect to $\ln(m)$ at fixed temperature gives the key result
\begin{equation}
%\boxed{
\frac{1}{k_{\mathrm{B}}T^2}
\frac{\partial E_{\rm a}}{\partial \ln m}
=
\frac{\partial b}{\partial T}.
%}
\label{eq:mainresult}
\end{equation}

Consequently we can conclude that, if $b$ is independent of temperature, then $E_{\rm a}$ is independent of mass; while if $b$ increases or decreases with temperature, the activation energy will scale in the same direction with mass.

\end{document}